\documentclass[prl,a4paper,twocolumn,superscriptaddress,nobalancelastpage]{revtex4-2}
\usepackage{amsmath,amsfonts,amsthm,graphicx,bm}

\newcommand{\ket}[1]{\lvert#1\rangle}
\newcommand{\bra}[1]{\langle#1\rvert}
\newcommand{\gras}[1]{\bold{#1}}
\newtheorem{theorem}{Theorem}

\usepackage{color}

\begin{document}

\title{Projected Entangled Pair States: Fundamental analytical and
numerical limitations}

\author{G. Scarpa}
\affiliation{\mbox{Dpto.\ An\'alisis Matem\'atico y Matem\'atica Aplicada, Universidad Complutense de Madrid, 28040 Madrid, Spain}}
\affiliation{\mbox{Instituto de Ciencias Matem\'aticas, 
Campus Cantoblanco UAM, C/ Nicol\'as Cabrera, 13-15, 28049 Madrid, Spain}}
\affiliation{Universidad Polit\'ecnica de Madrid, 
Escuela T\'ecnica Superior de Ingeniería de Sistemas Inform\'aticos, C/ Alan Turing 
s/n, 28031 Madrid, Spain}
\author{A. Moln\'ar}
\affiliation{\mbox{Dpto.\ An\'alisis Matem\'atico y Matem\'atica Aplicada, Universidad Complutense de Madrid, 28040 Madrid, Spain}}
\affiliation{\mbox{Instituto de Ciencias Matem\'aticas, 
Campus Cantoblanco UAM, C/ Nicol\'as Cabrera, 13-15, 28049 Madrid, Spain}}
\affiliation{Max-Planck-Institute for Quantum Optics, Hans-Kopfermann-Str. 1, 85748 Garching, Germany}
\affiliation{Munich Center for Quantum Science and Technology, Schellingstr.~4, 80799 M\"unchen, Germany}
\author{Y. Ge}
\affiliation{Max-Planck-Institute for Quantum Optics, Hans-Kopfermann-Str. 1, 85748 Garching, Germany}
\affiliation{Munich Center for Quantum Science and Technology, Schellingstr.~4, 80799 M\"unchen, Germany}
\author{J. J. Garc\'ia-Ripoll}
\affiliation{Instituto de F\'isica Fundamental IFF-CSIC, Calle Serrano 113b, 28006 Madrid, Spain}
\author{N. Schuch}
\affiliation{Max-Planck-Institute for Quantum Optics, Hans-Kopfermann-Str. 1, 85748 Garching, Germany}
\affiliation{Munich Center for Quantum Science and Technology, Schellingstr.~4, 80799 M\"unchen, Germany}
\affiliation{University of Vienna, Department of Mathematics,
Oskar-Morgenstern-Platz 1, 1090 Wien, Austria}
\affiliation{University of Vienna, Department of Physics,
Boltzmanngasse 5, 1090 Wien, Austria}
\author{D. P\'erez-Garc\'ia}
\affiliation{\mbox{Dpto.\ An\'alisis Matem\'atico y Matem\'atica Aplicada, Universidad Complutense de Madrid, 28040 Madrid, Spain}}
\affiliation{\mbox{Instituto de Ciencias Matem\'aticas, 
Campus Cantoblanco UAM, C/ Nicol\'as Cabrera, 13-15, 28049 Madrid, Spain}}
\author{S. Iblisdir}
\affiliation{\mbox{Dpto.\ An\'alisis Matem\'atico y Matem\'atica Aplicada, Universidad Complutense de Madrid, 28040 Madrid, Spain}}
\affiliation{\mbox{Instituto de Ciencias Matem\'aticas, 
Campus Cantoblanco UAM, C/ Nicol\'as Cabrera, 13-15, 28049 Madrid, Spain}}
\affiliation{Departament de Física Quàntica i Astronomia \& 
Institut de Ciències del Cosmos, Universitat de Barcelona, 08028 Barcelona, Spain}

\begin{abstract}
Matrix Product States (MPS) and Projected Entangled Pair States (PEPS) are
powerful analytical and numerical tools to assess quantum many-body
systems in one and higher dimensions, respectively. While MPS are
comprehensively understood, in PEPS fundamental questions, relevant 
analytically as well as numerically, remain open, such as how to encode
symmetries in full generality, or how to stabilize numerical methods using
canonical forms.  Here, we show that these key problems, as well as a number
of related questions, are algorithmically undecidable, that is, they
cannot be fully resolved in a systematic way.  Our work
thereby exposes fundamental limitations to a full and unbiased
understanding of quantum many-body systems using PEPS.
\end{abstract}

\maketitle

Matrix Product States (MPS) have proven highly successful in the study of
interacting of one-dimensional (1D) quantum systems.  The explanation of
the Density Matrix Renormalization Group method -- the method of choice
for the simulation of 1D systems -- as a variational method over the
manifold of MPS has triggered the development of a plethora of new
methods, such as for the simulation of time evolution, excitations,
or thermal states, as well as generalizations to higher dimensions using
Projected Entangled Pair States
(PEPS)~\cite{schollwoeck:rmp,schollwoeck:review-annphys,orus:tn-review,bridgeman:interpretive-dance}.
As their analytical understanding progressed, MPS were also exploited for
analytical studies, and have in particular led to a full and rigorous
classification of entangled phases under symmetries (``SPT phases'') in 1D~\cite{pollmann:symprot-1d,chen:1d-phases-rg,schuch:mps-phases}.

At first sight, this success is rooted in the fact that MPS provide a
faithful approximation to low-energy states of physical systems by 
capturing their entanglement
structure~\cite{verstraete:faithfully,hastings:arealaw,arad:rg-algorithms-and-area-laws}.
Nonetheless, a number of seemingly technical developments were central in
turning this basic idea into the
powerful numerical and analytical framework it is today; remarkably,
those insights were later found to be intimately related to the
fundamental structure of MPS.
First, in order to render the variational optimization stable, the use of
suitable \emph{canonical forms} is needed, which e.g.\ remove
singularities in the normalization or redundant degrees of
freedom~\cite{schollwoeck:review-annphys,haegeman:mps-ansatz-excitations}.
Second, in simulating systems with symmetries, those symmetries are often
explicitly encoded: This speeds up the method, allows to reach
significantly higher accuracies, ensures that the variational state
is perfectly symmetric, and provides means to directly extract
signatures of SPT order~\cite{schollwoeck:review-annphys,singh:tns-decomp-symmetry,bauer:abelian-sym-in-peps,weichselbaum:nonabelian-syms-in-tn,haegeman:1d-spt-orderparameter,pollmann:spt-detection-1d}.
Crucially, the possibility to numerically utilize symmetries hinges on the
fact that in MPS, any global symmetry of a wavefunction can be implemented
locally, that is, at the level of the individual tensor which describes a
single
site~\cite{perez-garcia:mps-reps,cirac:mpdo-rgfp,delascuevas:fund-thm-periodic}.

Concurrently, it was understood that the local encoding of symmetries in
MPS makes them an extraordinarily powerful analytical tool. Most
importantly, the classification of all such encodings has provided us with
a comprehensive classification of all phases in 1D under
symmetries~\cite{pollmann:symprot-1d,chen:1d-phases-rg,schuch:mps-phases}.  This
success relies on two crucial points: First, the exhaustive knowledge of
all ways in which any given symmetry can act in MPS, and second, the
ability to interpolate between any two MPS in the same phase such that along
the whole interpolation, there is a smooth, gapped parent Hamiltonian.

Both numerical and analytical uses of symmetries hinge on a point
of utmost importance: We must be absolutely certain that we  consider
all the possible ways in which symmetries of a 1D system can be
encoded locally in the MPS. Otherwise, any classification would
risk missing out on part of the possible phases, and correspondingly,
numerical methods would be biased and possibly even unable to
capture certain phases.
Fortunately, a full characterization of all symmetry realizations in MPS is
indeed known: It follows from the ``fundamental theorem of MPS,''
which fully characterizes how two different MPS representing
the same state -- such as obtained by acting with a symmetry -- are
related, and which in turn is intimately connected with the aforementioned
canonical forms relevant for numerical
stability~\cite{schollwoeck:review-annphys,perez-garcia:mps-reps,cirac:mpdo-rgfp,delascuevas:fund-thm-periodic}.

PEPS, the 2D analogue of MPS, similarly
form a faithful ansatz for low-energy states in
2D~\cite{hastings:locally,molnar:thermal-peps}.  Numerical PEPS
algorithms are progressing rapidly, even though canonical forms analogous to
1D are
lacking~\cite{orus:tn-review,corboz:iPEPS-CTM,vanderstraeten:iPEPS-gradient}.
Given the increased complexity in 2D, making use of symmetries is even
more important. Generally, this is done using a straightforward
generalization from 1D, where symmetries act on all entanglement degrees
of freedom
independently~\cite{singh:tns-decomp-symmetry,bauer:abelian-sym-in-peps,weichselbaum:nonabelian-syms-in-tn,jiang:sym-peps-phases}.
For long, this was the only symmetry realization considered, and thus, it
came as a big surprise when Chen, Liu, and Wen devised a model where the
symmetry, though local, acted in a correlated way on the entanglement
degrees of freedom, and showed that this constituted a non-trivial SPT
phase in
2D~\cite{chen:2d-spt-phases-peps-ghz,chen:spt-order-and-cohomology}.
Subsequently, such correlated symmetry actions have been found to also
underlie topologically ordered
phases~\cite{buerschaper:twisted-injectivity,sahinoglu:mpo-injectivity,bultinck:mpo-anyons}.
This contested how to encode symmetries in PEPS in the most general
way -- a question which is not only central to a comprehensive
classification of phases, but yet again equally for the use of symmetries
in numerical
simulations, such as to guarantee an unbiased approach. Thus, an analogous
``fundamental theorem of PEPS'', which elucidates the most general way to
realize symmetries in PEPS, is highly desirable. 

In this paper,  we show that such a result cannot exist: It is impossible
to fully characterize all the ways in which symmetries can be realized in
PEPS -- whatever list of possible realizations one has, it can never be
complete.  The reason is that, as we show, the question whether a
PEPS has a certain symmetry is undecidable, that is, there is no algorithm
which, given a PEPS tensor, can ever decide whether the family of states
generated by that tensor will be symmetric. In particular, this rules out
the possibility to have a list of possible symmetry realizations which can
be systematically checked. This implies that we cannot just use PEPS for
the classification of all possible phases, as we would risk losing
certain symmetry realizations: For the very least, we would have to impose
additional structure, such as to a physically motivated
subclass of PEPS, but it could just as well be entirely
impossible. Similarly, this implies that in numerical study, imposing
symmetries locally through the tensor will always rule out certain symmetric states, even if we try to
include as many symmetry realizations as possible. As a
consequence, our no-go result implies the impossibility of a ``fundamental
theorem of PEPS''.

A key primitive in our argument is the problem of assessing if a PEPS is
normalizable -- that is, given a PEPS tensor, does it describe a state
with non-zero norm?  This problem in fact is already of interest on its
own: A main reason for using canonical forms in numerical simulations is to
avoid divergences in the normalization of the state. In PEPS simulations, 
such ill-conditioned behavior which could be tied to convergence of the
method to non-normalizable states has indeed been
observed~\cite{xie:pess,christandl:rvb-limit}.  As
we show, this problem by
itself is undecidable, that is, there is no algorithm which can ensure
that a given PEPS tensor gives rise to properly normalizable states,
thereby avoiding singularities and stabilizing numerical simulations.
And finally, the undecidability of this key primitive has also another
consequence, relating to the classification of phases using PEPS: One part
of such a classification is the ability to connect two PEPS in the same
phase along a smooth and gapped path~\cite{schuch:mps-phases}. However, we show that generally, the
gap of such a parent Hamiltonian of a PEPS is undecidable as well.  While
this does not rule out suitable interpolations -- they have to be
constructed in a way which circumvents those cases -- it exposes yet
another limitation on the use of PEPS in a full classification of phases.

\begin{figure}[t]
\includegraphics[width=\columnwidth]{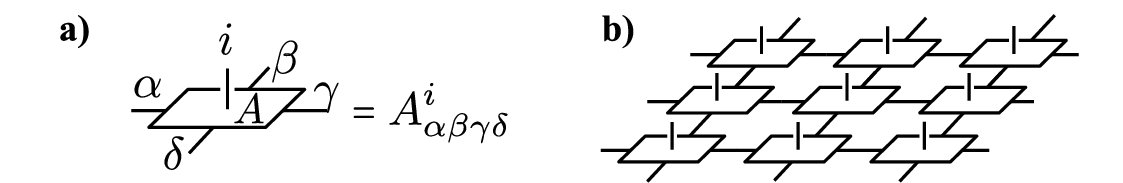}
\caption{PEPS construction: A five-index tensor (a) is used to describe a
wavefunction on an infinite lattice by contracting the virtual indices
$\alpha,\beta,\dots$ as indicated.}
\label{fig:peps-construction}
\end{figure}

Let us start by reviewing the formalism of PEPS.  A translational
invariant PEPS (iPEPS) is characterized by a $5$-index tensor $A\equiv
A^i_{\alpha\beta\gamma\delta}$ with \emph{physical index} $i$ and
\emph{virtual} or \emph{entanglement indices} $\alpha,\dots,\delta$. This
tensor generates a family of states on systems of arbitrary size
$N_h\times N_v$ (and thus asymptotically on the infinite plane) as
follows: We arrange the tensor $A$ in an $N_h\times N_v$ grid and contract (that is,
identify and sum over) adjacent indices, indicated by connected lines in
Fig.~\ref{fig:peps-construction}b, and terminate the virtual indices at the
boundary with some boundary condition.  The resulting object now
only depends on the physical indices $\bm i
=(i_{1,1},\dots,i_{N_h,N_v})$, and we denote it 
 by 
$\mathcal C_A(\bm i)$. It
defines a quantum state -- the
PEPS -- by virtue of 
$\ket{\Psi_{N_h,N_v}}=\sum_{\bm i}\mathcal C_A(\bm i)\ket{\bm i}$.
In principle, $\mathcal C$ still depends on the boundary conditions.
However, for a well-defined thermodynamic limit both in numerical
simulations and in a rigorous mathematical sense, we must demand that
properties in the bulk become independent of the chosen boundary
condition, except for possibly selecting a symmetry broken sector.

\begin{figure}[b]
\includegraphics[width=\columnwidth]{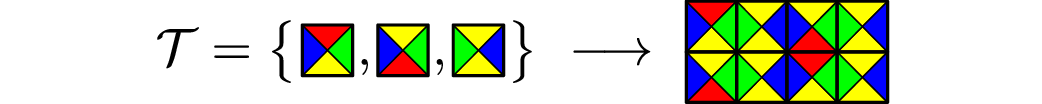}
\caption{The tiling problem: Can a set of tiles  tile the
plane? 
}
\label{fig:tiling}
\end{figure}

A key tool in our work will be the \emph{tiling
problem}~\cite{boerger:decision-problem-book}, Fig.~\ref{fig:tiling}. It is specified by
a set of square tiles $\mathcal T=\{t_i\}_{i=1}^T$, which are colored in a way where each edge
$e=l,u,r,d$ is assigned a color~$c_e$, $t\equiv(c_l,c_u,c_r,c_d)$.
  The tiling problem now asks to cover a given
region with those tiles (without rotating), such that the colors of two
adjacent tiles agree on shared edges.
Tiling problems can be mapped to a tensor network in a natural way: 
To this end, choose the dimension of the physical index equal to the
number of allowed tiles $|\mathcal T|$, and define a tensor
\begin{equation}
\label{eq:A-from-tile}
A^{t}_{\alpha_l\alpha_u\alpha_r\alpha_d} = 1
\mbox{\quad if\ }
t = (\alpha_l,\alpha_u,\alpha_r,\alpha_d)\ ,
\end{equation}
and zero otherwise.

What does this construction imply for the PEPS
$\ket{\Psi_{\mathrm{tile}}}$ constructed from $A$ on a patch $N_h\times
N_v$?
Consider two adjacent tensors
$A^{t}_{\alpha_l\alpha_u\alpha_r\alpha_d}$ and 
$A^{t'}_{\beta_l\beta_u\beta_r\beta_d}$ with tiles 
$t=(c_l,c_u,c_d,c_r)$ and 
$t'=(c'_l,c'_u,c'_d,c'_r)$,
where the former is
to the left of the latter. 
Contracting the tensors means setting $\alpha_r=\beta_l$ and
summing.  But since $\alpha_r=c_r$ and $\beta_l=c_l'$, this implies
$c_r=c_l'$, summed over all possible values.  Thus, the adjacent tile
colors of the tensors must match, and can take any allowed
value -- by construction, the
tensors automatically enforce the tiling rules.  It is now easy to see
that this pattern persists as one continues to contract indices.
Specifically, if the coloring $\bm t=(t_{1,1},\dots,t_{N_h,N_v})$ is
inconsistent, then 
$\mathcal C_A(\bm t)=0$, and otherwise,
$\mathcal C_A(\bm t)=\langle \bm\alpha\ket{b}$, where
$\bm\alpha$ is the boundary coloring of $\bm t$, and $\ket{b}$ the
boundary condition imposed.

A second way to map tiling problems to PEPS is to choose
$\tilde A^i_{\alpha_l\alpha_u\alpha_r\alpha_d}=1$ if $i=0$ and
$(\alpha_l,\alpha_u,\alpha_r,\alpha_d)\in\mathcal T$, and zero otherwise:
This construction yields a PEPS description $\ket{\tilde\Psi_{\mathrm{tile}}}$
of a product state $\ket{0}^{\otimes N}$ (or, alternatively, a tensor
network without physical indices), where the normalization equals the number of
all consistent tilings; again, if there is no consistent tiling, the norm
of this PEPS is zero.  Clearly, $\ket0$ can be replaced with any other
state.

A key result on tiling problems relates to the
following: Given a set of tiles, is it possible to use them to tile the
infinite plane -- that is, to tile regions of arbitrary size $N_h\times
N_v$, if we don't impose boundary conditions -- or is there a size above which no allowed tiling exists?
It has been shown that this problem is (algorithmically) undecidable, that
is, there is no algorithm which runs in finite time (no matter how slow)
which will solve this question for an arbitrary set of allowed
tiles~\cite{berger:tiling-undecidable,boerger:decision-problem-book}. Let
us now apply this result in the light of our tiling to PEPS mapping:
Since $\mathcal C_A(\bm t)$ can be non-zero only
for allowed tilings, the impossibility to tile the plane implies that there
is a size $N_h\times N_v$ above which the PEPS defined by the tensor $A$,
Eq.~\eqref{eq:A-from-tile}, is identically zero (that is,
non-normalizable), \emph{regardless} of the choice of boundary conditions.
On the other hand, if a valid tiling exists, the state will be non-zero
for a suitable boundary condition, such as the  uniform superposition of
all colors.
 We thus find that the following problem is undecidable:
\\[1ex] 
\emph{\textbf{Undecidable problem 1: PEPS zero testing.} Given a tensor
$A$, is there a choice of boundary conditions such that the PEPS
constructed from $A$ is non-zero for all system sizes?}
\\[1ex] 
Note that the boundary conditions are allowed to depend on the system
size.  Importantly, the non-existence of such boundary conditions in
particular implies that the PEPS described by $A$ is ill-defined 
 in the thermodynamic limit.  
Using the second
construction above, this result even holds for PEPS without a physical
index (or non-trivial PEPS representations of product states).

As explained, this problem is of considerable interest by itself to
prevent PEPS algorithms to run into singular points; our result shows
that there is no algorithm which will circumvent this problem in the
general case.  However, it will moreover serve us as an elementary
building block to assess the difficulty of central problems in the field
of PEPS, most importantly the realization of symmetries and the existence
of a fundamental theorem.

We start by considering the problem of how to implement symmetries in PEPS
in full generality.  As outlined above, a comprehensive understanding is
essential both for a complete classification of phases and for a fair and
unbiased implementation of symmetries in numerical simulations.  
First off, we need to clarify what it means for an
iPEPS to be symmetric. To start with, let us consider an
on-site symmetry $\mathcal S_g = U_g^{\otimes N}$, $g\in G$, which is
unbroken.  A PEPS tensor provides a ``good'' description of an infinite
system if it yields a well-defined limit as we increase the system size,
independent of the boundary condition we choose. This is both true in
numerical simulations, where we require well-defined corner transfer
matrix (CTM) or iMPS fixed
points~\cite{orus:tn-review,corboz:iPEPS-CTM,vanderstraeten:iPEPS-gradient},
as well as the correct formal definition in mathematical physics.
Specifically, for a symmetric PEPS, this means that the reduced density
matrix $\rho_R$ of any region $R$, in the limit where the boundaries are
far away, is unique and thus invariant under the symmetry, $\mathcal
S_g\rho_R\mathcal S_g^\dagger=\rho_R$.  This characterization  generalizes
immediately to other symmetries, such as reflection, time-reversal, or
translation (in the last case, $\rho_R$ has to be invariant under
translation of $R$); and can be generalized to broken symmetries, where
$\rho_R$ can depend on the boundary condition only to the extent that the
boundary selects a symmetry broken sector.

Now consider an arbitrary PEPS $\ket{\Theta_\mathrm{sym}}$ with tensors
$B^i_{\alpha_l\alpha_u\alpha_r\alpha_d}$ which is invariant under the
symmetry $\mathcal S_g$ in the sense above.  In addition, take a product
state which is \emph{not} invariant under $\mathcal S_g$ for some $g$
(such a state always exists, possibly not translation invariant if
point/space group symmetries are included), and construct a PEPS
description $\ket{\tilde\Psi_\mathrm{tile}}$ (with tensors $\tilde A^i$)
of this product state, following the mapping from the tiling problem
described above, for which the question
$\ket{\tilde{\Psi}_\mathrm{tile}}\stackrel{?}{=}0$ is
undecidable. 
We can now combine these two PEPS into a single PEPS 
\[
\ket{\Phi_\mathrm{sym?}} = 
\ket{\tilde\Psi_\mathrm{tile}} + \ket{\Theta_\mathrm{sym}} \ ;
\]
its PEPS tensor $C$ is obtained as the direct sum of $\tilde A^i$ and
$B^i$ (that is, the virtual space is the direct sum of the two virtual
spaces, $C^i=\tilde A^i$ exactly if all virtual indices are in the
$\tilde A$ part
of the direct sum, $C^i=B^i$ exactly if all are in the $B$ part, and
$C^i=0$ otherwise).
 Let us now
consider the two cases which are undecidable to distinguish: First, 
there is a valid tiling, and thus $\ket{\tilde{\Psi}_\mathrm{tile}}\ne0$
for some suitable boundary condition for any system size -- in that case,
the reduced density matrix $\rho_R$ of $\ket{\Phi_\mathrm{sym?}}$ will
depend on the boundary conditions, and in particular, there exist boundary
conditions supported in the $\tilde A^i$ sector 
for which $\rho_R$ will not be symmetric.  Second, there is no
valid tiling of the plane, and thus
$\ket{\tilde{\Psi}_\mathrm{tile}}\equiv0$ for large enough systems -- then, for any boundary condition, $\rho_R$ will only have contributions
from $\ket{\Theta_\mathrm{sym}}$, and thus be symmetric.  We therefore
find that the following problem is undecidable:
\\[1ex] 
\emph{\textbf{Undecidable problem 2: Symmetries in PEPS.} Given a tensor $C$ and a symmetry
$\mathcal S_g$, $g\in G$, is the translational invariant PEPS constructed from $C$
invariant under $\mathcal S_g$ in the thermodynamic limit?}
\\[1ex]
Note that this result holds for any symmetry representation, and thus
e.g.\ already for spin-$\tfrac12$ systems.

This undecidability statement has profound implications for the
classification of symmetry encodings in PEPS.  In particular, it tells us
that there cannot be a comprehensive list of all possible ways to encode
symmetries in PEPS which can be checked by any algorithm (regardless how
inefficient), since otherwise, there would be an algorithm deciding the
presence or absence of symmetries in a given PEPS. Therefore, any
classification of symmetries in PEPS, and thus any classification of
phases under symmetries, must rely on some assumptions about the way in
which the symmetry is encoded, or in some other way restrict to a subclass
of PEPS with additional structure.

A direct consequence is the impossibility to have a ``fundamental
theorem of PEPS'' which relates two different PEPS representations of the
same state.
\\[1ex] 
\emph{\textbf{Undecidable problem 3: Fundamental Theorem for PEPS.} Given
tensors $A^i$ and $B^i$, decide if they describe the same PEPS for any
system size.}
\\[1ex] 
Here, ``describe the same PEPS'' is meant in the same way as for
symmetries, namely, that the reduced density matrices of the two states
become indistinguishable as the boundaries are taken to infinity.  The
proof is immediate since by choosing $B^i = \sum (U_g)_{ij} A^j$, we see
that the case of deciding symmetries (Undecidable problem 2) is a special
case. Importantly, this implies that there is no algorithm to bring
PEPS into a canonical form from which we can algorithmically decide
whether they describe the same state, let alone find a local gauge
transformation relating $A^i$ and $B^i$.

Finally, we can employ our construction to show
that verifying whether a PEPS parent Hamiltonian is gapped is an
undecidable problem. As discussed in the introduction, PEPS come naturally
with parent Hamiltonians%
~\cite{perez-garcia:parent-ham-2d,schuch:peps-sym}, and
in 1D, the fact that those Hamiltonians are uniformly gapped is a crucial
ingredient in the classification of phases, as it allows to connect two
states in the same phase through a gapped path of MPS~\cite{schuch:mps-phases}. However, in 2D, the
corresponding problem is undecidable:
\\[1ex]
\emph{\textbf{Undecidable problem 4: Gap of parent Hamiltonians.}
For the parent Hamiltonian $H_N$ of a PEPS with open boundaries on an $N\times
N$ patch, it is undecidable to distinguish between the following two
cases: 
(i) $H$ is gapped -- specifically, there exists an $N_0$
such that for all $N\ge N_0$, $H$ has a spectral gap $\Delta\ge1$ above
the ground state;
(ii) $H$ is gapless -- specifically, the spectrum of $H$ converges to
$[0;\infty)$. 
}
\\[1ex]
The proof proceeds by constructing a PEPS which is a superposition of a
PEPS with a gapped parent Hamiltonian and a PEPS with a gapless parent
Hamiltonian, coupled to a tiling PEPS without physical indices.  This way,
whenever there does not exist a tiling, we obtain the first state and thus
a gapped Hamiltonian, while otherwise, the other state appears and gives
rise to a gapless parent Hamiltonian (see Supplemental Material for
details~\cite{suppmat}).  Note that this strengthens the
existing undecidability result for spectral
gaps~\cite{cubitt:undecidable-gap-short}, as it shows it even
holds when restricting to PEPS parent Hamiltonians.

Finally, the mapping between tilings and PEPS, applied to
finite systems with fixed boundary conditions, allows for straightforward
proofs of the \textsf{NP}-hardness of certain problems, such as finding
ground states and zero testing, presented in the Supplemental
Material~\cite{suppmat}.

In conclusion, we have studied the difficulty of some central problems in
the field of PEPS, relevant both for their use as an analytical framework
and as a powerful numerical tool.  By establishing a mapping between
tiling problems and PEPS, combined with the existence of
undecidable
tiling problems, we were able to establish undecidability of a range
of these problems. Specifically, we could show that the problem of deciding
whether a PEPS is invariant under a given symmetry, as well as the problem
of deciding whether two PEPS tensors represent the same state, are
undecidable. In particular, this implies the impossibility to
succinctly enumerate all ways in which a symmetry can be encoded locally
in the PEPS tensor, and thus exposes fundamental limits to the possiblity
of performing unbiased analytical and numerical analysis of PEPS with
symmetries. As a technical result with implications on its own right, we
could moreover show that the problem of deciding whether a PEPS tensor
actually describes a physical (that is, non-zero and thus normalizable)
state is undecidable as well; and finally, we have shown how this implies
undecidability of a spectral gap in the case of PEPS parent Hamiltonians.
Let us note that these limitations, while severe, can potentially be
overcome by restricting to a physically relevant subclass of PEPS, such as
the states which appear in approximation proofs or numerical simulations,
and which likely carry additional structure which could potentially allow
to overcome those restrictions.

\begin{acknowledgments}
\emph{Acknowledgements.---}We thank Johannes Bausch for helpful comments on
the manuscript. 
This project has received funding from the European Research Council (ERC)
under the European Union's Horizon 2020 research and innovation programme
(grant agreements No 648913 and 636201). 
SI, DPG, and GS acknowledge financial support from
MINECO (grants MTM2014-54240-P and MTM2017-88385-P), CAM research
consortion QUITEMAD-CM (grant P2018/TCS-4342), and Severo Ochoa project
SEV-2015-556. 
SI also acknowledges funding by the Ministerio de Ciencias, Innovación y
Universidades, PGC2018-095862-B-C22 (Spain) and by  Quantum CAT (001-P-001644),
Generalitat de Catalunya (Spain).   
JJGR acknowledges financial support from MINECO project
FIS2015-70856-P, CAM research consortion QUITEMAD-CM Grant
No.~P2018/TCS-4342, MINECO research network FIS2016-81891-REDT. AM, YG,
and NS acknowledge support from the DFG (German Research Foundation) under
Germanys Excellence Strategy (EXC2111-390814868).
\end{acknowledgments}

\onecolumngrid
\begin{center}
\textbf{SUPPLEMENTAL MATERIAL}
\end{center}
\vspace*{3em}
\twocolumngrid

\section{I. \textsf{NP}-hardness of commuting local Hamiltonian and zero testing}

Here, we show that the \textsf{NP}-hardness of the tiling problem on
finite systems allows for simple \textsf{NP}-hardness proofs for the 
commuting local Hamiltonian problem, as well as -- if
combined with the mapping from tiling problems to PEPS -- for an
\textsf{NP}-hardness proof of PEPS zero testing on finite systems.

\subsection*{\textsf{NP}-hardness of tiling problems}

We start by introducing the Bounded Tiling problem.
\\[1ex]
\emph{\textbf{Bounded Tiling:} Fix a set of tiles. The Bounded Tiling
problem asks whether it is possible to tile a rectangle of size $N_h\times
N_v$ with given boundary conditions (i.e., colors). The Bounded Tiling
problem is \textsf{NP}-complete~\cite{EmdeBoas84}.
}
\subsection*{\textsf{NP}-hardness of commuting local Hamiltonians}

Bounded Tiling admits a formulation in terms of a classical
nearest-neighbor Hamiltonian. We consider a square lattice
$\Lambda$ with local Hilbert space $\mathcal{H}_{\rm tiling}= {\rm span}
\{\ket{t}\}$, where the $t$'s run on the tiles in the given tile set.  The
energy operator is \begin{equation}\label{eq:hbt}
H=\sum_{\langle p,p' \rangle} h_{p,p'}^{\rm tiling}+ \sum_{p
\in \partial \Lambda} h_p^{\partial}\ .
\end{equation}
Here, $h^{\rm tiling}_{p p'}$ is the (classical) Hamiltonian that gives energy $1$ to invalid tiling configurations. That is, $h^{\rm tiling}_{p p'}=\sum \ket{t}\bra{t}_p\otimes \ket{t'}\bra{t'}_{p'}$ where the sum runs on all invalid configurations of neighboring tiles. Note that the interaction is translational invariant but not rotational invariant. Similarly, $h_p^\partial$ acts on any plaquette $p$ at the boundary, and
enforces the boundary conditions by assigning energy $0$ to tiles matching
the boundary conditions, and $1$ otherwise.

It is now clear that the Hamiltonian $H$ constructed this way has ground
state energy $0$ if and only if an allowed tiling exists, that is, it
corresponds to a ``yes'' instance of the Bounded Tiling problem.
We thus obtain the following hardness result:
\\[1ex]
\emph{\textbf{Translational invariant commuting $2$-local Hamiltonian
problem on square lattice:} Given a translational invariant 2-body
Hamiltonian on a square lattice of size $N_h\times N_v$,
together with a set of boundary terms, and where all terms commute,
deciding whether the ground state minimizes the energy of each term
individually  is \textsf{NP}-complete.
}
\\[1ex]
Note that while a similar hardness result also follows e.g.\ from
Barahona's result that finding the ground state energy of a bilayer spin
glass is \textsf{NP}-hard \cite{barahona}, our construction additionally
yields a Hamiltonian which is translational invariant in the bulk.
Containment of the $2$-local commuting Hamiltonian problem in \textsf{NP},
also for non-classical commuting Hamiltonians, has been shown by Bravyi
and Vyalyi~\cite{bravyivyalyi}.

\subsection*{\textsf{NP}-hardness of PEPS zero-testing}

Just as we did in the main body the paper, we can also map any \textsf{NP}-hard
Bounded Tiling problem to a tensor network on an $N_h\times N_v$ patch with boundary
conditions.  The \textsf{NP}-hardness of the
question whether there exists a tiling then translates into testing
whether the resulting PEPS, with the given boundary conditions, is
non-zero.  We thus arrive at the following \textsf{NP}-hardness result:
\\[1ex]
\emph{\textbf{PEPS zero testing:}
The following problem is \textsf{NP}-hard: Given a PEPS tensor $A^i$ and boundary
conditions for a region $N_h\times N_v$, is the resulting PEPS on that
region non-zero?}
\\[1ex]
The \textsf{NP}-hardness of PEPS zero testing has previously been established in Ref.~\cite{ref7} through a reduction from the coloring problem; for the 
special case of tensors with non-negative entries (such as in our construction), \textsf{NP}-completeness was shown.  As compared to that construction, or a direct \textsf{NP}-hardness proof by encoding the
\textsf{NP} verifier circuit into a tensor network,
our construction has once more the advantage that 
it yields a tensor network which is
translational invariant in the bulk.

\section{II. Gap in the parent Hamiltonian}
The undecidability of the spectral gap for short-range Hamiltonians has
been established in \cite{cubitt:undecidable-gap-short}.
 We now show that even if we restrict to Hamiltonians that are parent Hamiltonians of a PEPS, undecidability still holds, at least in the case of a finite but unbounded local physical dimension. For that purpose, let us recall a standard procedure to associate a nearest neighbor parent hamiltonian with a PEPS described by a tensor $A$ \cite{perez-garcia:parent-ham-2d}.
 With any region $R$ of the lattice, we associate a linear map  
\[
\chi(A,R): \big( \mathbb{C}^D \big)^{\otimes |\partial R|} \to \big( \mathbb{C}^d \big)^{\otimes |R|}:
\ket{C} \to \sum_{i_R} \mathcal{C} [A_{i_R} C] \ket{i_R}.
\]
A parent Hamiltonian is any nearest neighbor Hamiltonian $H=\sum_{\langle p,p' \rangle} h_{p,p'}$, such that $h_{p,p'} \geq 0$, and such that
\begin{equation}\label{eq:parent-property}
\text{Ker} \; h_{p,p'}=\text{Im} \; \chi(A,p \cup p').
\end{equation}

The aim of this appendix is to prove the following

\begin{theorem}
For the parent Hamiltonian $H_N$ of a PEPS with open boundaries on an $N\times
N$ patch, it is undecidable to distinguish between the following two
cases: 
(i) $H$ is gapped -- specifically, there exists an $N_0$
such that for all $N\ge N_0$, $H$ has a spectral gap $\Delta\ge1$ above
the ground state;
(ii) $H$ is gapless -- specifically, the spectrum of $H$ converges to
$[0;\infty)$. 
\end{theorem}

For that, we need as ingredients the tensor $A$ in the main text giving rise to the PEPS $\ket{\Psi_{\rm tile}}$ and the tensor $B$ giving rise to the Ising PEPS $\ket{\Psi_{\rm Ising}}$ \cite{david:peps-ising}. Call $\mathcal{H}_{\rm tiling}$ and $\mathcal{H}_{\rm Ising}$ to their corresponding local physical Hilbert spaces, that is, $\mathcal{H}_{\rm Ising}$ is 2-dimensional and $\mathcal{H}_{\rm tiling}= \mathrm{span}\{\ket{t}\}$ where the $t$'s are the tiles of a given tiling problem. Let us finally assume that $\mathcal{H}_0={\rm span}\{\ket{0}\}$ is orthogonal to the space $\mathcal{H}_{\rm tiling}\otimes\mathcal{H}_{\rm Ising}$

We will be interested in the state $\ket{\Psi}=\ket{\Psi_0}+ \ket{\Psi_{\rm tile}} \otimes \ket{\Psi_{\rm Ising}}$, where $\ket{\Psi_0}$ is just the all--$\ket{0}$ product state, which is a PEPS with bond dimension $1$. Let us 
denote its defining tensor by $P$.

We will take $\mathcal{H}= \mathcal{H}_0\oplus \mathcal{H}_1$, with $\mathcal{H}_1= \mathcal{H}_{\rm tiling}\otimes\mathcal{H}_{\rm Ising}$
as our local Hilbert space. $\Pi^0$ and $\Pi^1$ will denote the projections onto the corresponding sector. 
$\ket{\Psi}$ is clearly a PEPS with local Hilbert space $\mathcal{H}$ if we use the direct sum construction of the main text. That is, the tensor associated to $\ket{\Psi}$ is just $P\oplus A\otimes B$, where the direct sum and tensor product are on each of the indices.
Define the following nearest neighbor interaction:
\begin{align*}
h_{p,p'} =& 
\Pi^0_{p} \otimes  \Pi^1_{p'} +\Pi^1_{p} \otimes  \Pi^0_{p'} \\
&+h^{\rm tiling}_{pp'} \otimes \gras{1}_{p p'}^{\rm Ising}+\gras{1}^{\rm tiling}_{p p'} \otimes h^{\rm Ising}_{p p'}\ ,
\end{align*}
where $\gras{1}$ is the identity in the corresponding Hilbert space, $h^{\rm Ising}_{pp'} $ is the parent interaction of the Ising PEPS and $h^{\rm tiling}_{p p'}$ is the classical interaction defined in \eqref{eq:hbt}. Note that the interactions in the second row are defined in $\mathcal{H}_1\otimes \mathcal{H}_1$. We assume them defined by $0$ in the other sectors.

The intuition behind this choice is that the terms in the first row force the ground state to be totally contained in one of the two sectors of the local Hilbert space, while the terms in the second row enforce the appropriate ground states in each of the sectors. 

Using that all the terms in the Hamiltonian commute it is easy to see that 
\begin{equation}\label{eq:undec-parent}
\text{ker} \; h_{p,p'}= {\rm span}\{\ket{00}\}\oplus({\rm ker}\,  h^{\rm tiling}_{p p'} )\otimes 
({\rm ker}\,  h^{\rm Ising}_{p p'})\ ,
\end{equation}
and that if $H_N$ is the Hamiltonian obtained from the interactions $h_{p,p'}$ on a $N\times N$ patch with open boundary conditions, then
$${\rm spec}(H_N)= \{0\} \cup\left( {\rm spec}(H_N^{\rm tiling}) +  {\rm spec}(H_N^{\rm Ising})\right).$$ 

This is explained in full detail in \cite{cubitt:undecidable-gap-short}.

Moreover, $h_{p,p'}$ is a parent Hamiltonian for the PEPS $\ket{\Psi}$ associated to the tensor $F=P\oplus A\otimes B$, since trivially $\text{Im} \; \chi(F,p \cup p')$ coincides with the expression \eqref{eq:undec-parent}.

Using that ${\rm spec}(H_N^{\rm Ising})$ converges 
to $[0,+\infty)$ 
as $N\rightarrow \infty$,
and that it is undecidable to distinguish $0\in {\rm spec}(H_N^{\rm tiling})$ from ${\rm spec}(H_N^{\rm tiling})\subset[1,\infty)$ (since both options correspond respectively to the existence or absence of a valid $N\times N$ tiling), we obtain the result. 

Finally, note that if a nearest-neighbor parent Hamiltonian of a PEPS is gapped, then all of them are. Hence, in the above theorem, one can always take as a parent Hamiltonian the one where $h_{p,p'}$ is the projector onto $(\text{Im} \; \chi(F,p \cup p'))^\perp$.

\end{document}